\documentclass[doublecol]{epl2} 
\usepackage{amssymb}   
\usepackage{amsmath}

\title{Unconventional charge and spin dependent transport properties
  of a graphene nanoribbon with line-disorder} \shorttitle{Graphene
  nanoribbon with line-disorder} 

\author{Sudin Ganguly\inst{1} \and Saurabh Basu\inst{1} \and Santanu
  K. Maiti\inst{2}} \shortauthor{S. Ganguly \etal}

\institute{ \inst{1} Department of Physics, Indian Institute of
  Technology Guwahati, Guwahati-781 039, Assam, India
  
  \inst{2} Physics and Applied Mathematics Unit, Indian Statistical
Institute, 203 Barrackpore Trunk Road, Kolkata-700 108, India
}

\pacs{72.80.Vp}{Electronic transport in graphene}
\pacs{72.25.-b}{Spin polarized transport}
\pacs{74.62.En}{Effects of disorder} 

\abstract{ Electronic transport
  with a line (or a few lines) of Anderson type disorder in a zigzag
  graphene nanoribbon is investigated in presence of Rashba spin-orbit
  interaction. Such line disorders give rise to peculiar behavior in
  both charge as well as spin-polarized transmission in the following
  sense. In the weak disorder regime, the charge transport data show
  Anderson localization up to a certain disorder strength, beyond
  which the extended states emerge and start dominating over the
  localized states. These results are the hallmark signature of a
  selectively disordered (as opposed to bulk disorder) graphene
  nanoribbon. However, the spin-polarized transport shows a completely
  contradicting behavior. Further, the structural symmetries are shown
  to have an important role in the spintronic properties of the
  nanoribbons. Moreover, the edge-disorder scenario (disorder
  selectively placed at the edges) seems to hold promise for the
  spin-filter and switching device applications. }

\begin{document}

\maketitle

\section{\label{intro}Introduction}

Since 2004, graphene~\cite{novo} has been attracted wide attention in
both theoretical and experimental research community due to its exotic
electronic and transport properties~\cite{neto}. Owing to some of
these properties, such as long spin-diffusion lengths (up to $\sim
100\,\mu$m)~\cite{luis,tombros, zomer,yang,han}, quasi-relativistic
band structure~\cite{novo,zhang}, unconventional quantum Hall
effect~\cite {novo,zhang,vp}, half metallicity~\cite{jun,lin} and high
carrier mobility~\cite{du,bolotin}, graphene was thought to be a
suitable candidate in spintronic applications. However, due to the
absence of a band gap~\cite{neto}, that possibility gets restricted.

The hurdle can be tackled by fabricating graphene into
quasi-one-dimensional ribbons where non-zero band gapes have been
found. The electronic properties of these graphene nanoribbons (GNR)
depend on the edge geometry~\cite{fujita,saito}. Based on the edge
structure, GNRs can have zigzag and armchair edges. Armchair GNRs
(AGNR) can be either metallic or semiconducting in nature depending
upon the width of the ribbon, whereas, zigzag GNRs (ZGNR) are always
metallic~\cite{fujita}. Due to the long spin-diffusion length, spin
relaxation time, and electron spin coherence
time~\cite{yazyev-prb,yazyev-prl,cantele}, GNRs are one of the most
promising candidates as spintronic device applications among the other
derivatives of graphene and have been studied extensively.

The presence of spin-orbit (SO) coupling, specifically the Rashba SO
coupling is the key factor in the spintronic
applications~\cite{spa1,spa2,spa3,spa4,spa5,spa6,spa7}. Though
  the strength of Rashba SO coupling in pristine graphene is very
  weak~\cite{gmitra-prb-2009} ($\sim 10\,\mu$eV), it can be enhanced
  by growing graphene layer on metallic substrates. Graphene grown on
  WSe$_2$ show Rashba coupling about $0.6\,$meV as predicted by Gmitra
  {\it et al}~\cite{gmitra-prb-2016}. Recent experimental observations
  showed that the strength of the Rashba SOC can be about $225\,$meV
  in epitaxial graphene layers grown on the Ni-surface~\cite{dedcov}
  and a giant Rashba SOC ($\sim$ $600\,$meV) from Pb intercalation at
  the graphene-Ir surface~\cite{calleja}.

The electrical properties of GNR can also be tuned by means of various
ways, such as chemical edge modifications~\cite{wang-prb-2007} or
chemical doping~\cite{ouyang,huang}, geometrical
deformation~\cite{tang-apl,bets-nano}, application of uniaxial
strain~\cite{sena,chang-jap}, and many more. Since the defects,
impurities or disorder are inevitable in graphene-based material, it
is important to study their effects on the spintronic properties of
GNRs. Different kinds of controllable defects such as Stone-Wales
defect~\cite{park-apl,ren-jap}, adatoms~\cite{weeks,sudin-mrx},
vacancies~\cite{yan-prb,sahin}, substitution~\cite{peres}, line
defects~\cite{lahiri-nano,lin-prb,okada,gunlycke,filho,santanu-jap-line},
and disorder~\cite{long-prl,wurm} may also alter the electrical
properties of GNR. However, in most of these references, the primary
goal was to tune the gap in the energy spectrum.

In this work, we shall be focusing on the spintronic properties of
GNR.  Moreover, Filho {\it et al.} showed in their work~\cite{filho}
that the introduction of a line of impurities can open up a gap in GNR
and thus the study of spintronic properties of GNRs can be
particularly of interest in this scenario. Another interesting
phenomenon studied by Zhong and Stocks~\cite{zhong-shell-dope} is that
the electron transport in shell-doped nanowire shows peculiar
behavior. The electron dynamics of shell-doped nanowire behaves
completely different from uniformly doped nanowires. There exists a
localization/quasi-delocalization transition, where specifically the
localization phenomenon dominates in the weak disorder regime,
  while it dampens in the strong disorder range. In this work,
motivated by this shell-doped scenario we introduce a line (or lines)
of impurities along the zigzag chains in ZGNR (see
Fig.~\ref{setup}). We shall demonstrate that with this kind of line
impurities, it is possible to tune the spin-transport
properties. Moreover, since the presence of line impurities destroys
the longitudinal mirror symmetry along the $x$ and $z$-axes of the
ZGNR, all the three components of the spin-polarized transmission
($P_x$, $P_y$ and $P_z$) will be finite, which was untrue in a
pristine ZGNR. Thus line-disordered ZGNR can serve as an efficient
spin-filter device.

In the present work, we explore different aspects of spin transport in
a line-disordered ZGNR in presence of Rashba SO interaction using
Landauer-B\"{u}ttiker formalism. We believe that such a study of the
spintronic properties has not been done for a line-disordered ZGNR so
far.

We organize the rest of the work as follows. In the next section, we
introduce the line-disordered ZGNR and the theoretical framework for
the total transmission and spin-polarized transmission. Based on the
theoretical framework, next we include an elaborate discussion of the
results where we have demonstrated the behavior of the three
components of the spin-polarized transmission, as a function of
different positions for the single line-disorder and also for
situations with multiple disorder lines. We end with a brief summary
of our results stating the highlights of our findings.

\section{\label{theory}Junction Setup and theoretical formulation}

\begin{figure}[ht]
\centering
\includegraphics[width=0.45\textwidth]{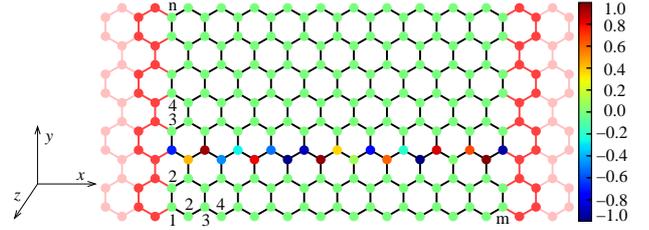} 
\caption{(Color online). Schematic illustration of a two-terminal ZGNR
  with a line-disorder. The disorder sites are denoted by random
  colors chosen from the color bar given at the right side of the
  geometry. The different colors indicate the value of the random
  on-site potentials at the carbon atoms, and are taken from a
  rectangular random distribution between $[-1:1]$. Rest of the carbon
  atoms have the same on-site potential (zero) denoted by the green
  color. The left and right leads are represented by the red color.}
\label{setup}
\end{figure}

Figure~\ref{setup} represents the schematic illustration of the model
quantum system, where a finite size ZGNR is coupled to two
semi-infinite pristine graphene leads with zigzag edges. The leads are
denoted by red color. A line of impurities denoted by different colors
is introduced along a single zigzag chain. The different colors denote
the random on-site potential and are picked up from the given color
bar. Apart from this zigzag line, the carbon atoms at all other sites
in the ZGNR are denoted by green color and their on-site potential is
fixed at zero as is seen from the color bar.
The Rashba SO interaction is assumed to be present in the central
scattering region only, while the leads are free from any kind of SO
interaction and disorder. Along the $x$-direction, the system has a
zigzag shape and along the $y$-direction it has the structure of an
armchair, and hence the system is conventionally defined as mZ-nA.

The tight-binding Hamiltonian modeled on a ZGNR in presence of Rashba
SO interaction can be written as~\cite{km1,km2},
\begin{equation}
H= \sum\limits_{i} \epsilon_i c_i^{\dagger} c_i -
t\sum\limits_{\langle ij\rangle}c_i^{\dagger} c_j +
i\alpha\sum\limits_{\langle ij\rangle}c_i^{\dagger} \left(
\vec{\sigma} \times {\bf\hat{d}}_{ij}\right)_z c_j
\label{h2}
\end{equation} 
where $\epsilon_i$ stands for the random on-site potential at the
$i$-th carbon atom chosen from a uniform rectangular distribution
($-W$ to $W$). $c_i^{\dagger}=\left(c_{i\uparrow}^{\dagger} \quad
c_{i\downarrow}^{\dagger}\right)$. $c_{i\sigma}^{\dagger}$
$(\sigma=\uparrow,\downarrow)$ is the creation operator of an electron
at site $i$ with spin $\sigma$. The second term is the
nearest-neighbor hopping (NNH) term. For simplicity, we assume that
the hopping strength between ordered-ordered, ordered-disordered and
disordered-disordered carbon atoms has the same value, and that is,
$t$. The third term is the nearest-neighbor Rashba term which
explicitly violates $z\rightarrow -z$ symmetry. $\vec{\sigma}$ denotes
the Pauli spin matrices and $\alpha$ is the strength of the Rashba SO
interaction. ${\bf\hat{d}}_{ij}$ is the unit vector that connects the
nearest-neighbor sites $i$ and $j$.

The total transmission coefficient, $T$ can be calculated
via~\cite{caroli,Fisher-Lee,dutta},
\begin{equation}
T = \text{Tr}\left[\Gamma_L {\cal G}_R
  \Gamma_R {\cal G}_A\right]
\label{trans-eq}
\end{equation}
where ${\cal G}_{R(A)}$ is the retarded (advance) Green's function.
$\Gamma_{L(R)}$ are the coupling matrices representing the coupling
between the central region and the left (right) lead. Also, the
spin-polarized transmission coefficient $P_\alpha$ can be calculated
from the relation~\cite{chang},
\begin{equation}
P_\alpha = \text{Tr}\left[\hat{\sigma}_\alpha\Gamma_L {\cal G}_R
  \Gamma_R {\cal G}_A\right]/T
\end{equation}
where, $\alpha=x,y,z$ and $\sigma$ denote the Pauli matrices, and $T$
as given in Eq.~\ref{trans-eq}.

\section{\label{randd}Numerical Results and discussion}

We set the hopping term $t=2.7\,$eV~{\cite{neto}}. All the energies are
measured in units of $t$. The strength of Rashba coupling is fixed at
a value given by $\alpha=0.1$, which is very close to the experimentally 
realized data~{\cite{dedcov}. The dimension of the scattering region in 
this work is taken as 241Z-40A. The widths of left and right leads are 
same as that of the central scattering region. For most of our numerical 
calculations, we have used KWANT~\cite{kwant}. 

We have essentially studied the behavior of total transmission $T$ and 
all the three components of the spin-polarized transmission $P_\alpha$ 
of a ZGNR with line-disorder in presence of Rashba SO interaction. All the 
results obtained below are averaged over $500$ distinct disordered 
configurations.

\subsection{Total transmission}
 
To begin with, we study the effect of a single line-disorder located
at the edges of the ZGNR, specifically, at the top edge or the bottom
edge.  Subsequently, we study the cases with a number of such disorder
lines, where half of the central scattering region is disordered and
compare the results with that of a bulk disordered ZGNR.

Figure~\ref{t}(a) shows the behavior of the transmission coefficient
as a function of the Fermi energy in presence of edge-disorder cases.
The line-disorder is located at either one of the edges of the ZGNR,
that is at the top (green line) or the bottom (black line) edge.  The
disorder strength is fixed at $W=0.5$. The transmission coefficient is
symmetric about $E=0$ and shows identical behavior for the two
different edge-disorder cases. Thus, the electronic charges do not
feel any difference whether the line-disorder is located at the top or
at the bottom edge of the sample. This is owing to the fact that the
elements of the $S$-matrix have certain symmetries owing to the
geometry of the sample, that is the reflection symmetry
$y\rightarrow-y$~\cite{kim-jap}. Further, the transmission coefficient
exhibits a peculiar behavior as a function of the disorder strength
$W$ as shown in Fig.~\ref{t}(b). For the Fermi energy fixed at a value
$E=-0.18$, corresponding to both the top and bottom edge-disorder
cases, $T$ shows similar behavior as already seen in
Fig.~\ref{t}(a). For the lower values of disorder, electrons tend to
localize, and as a result $T$ decreases. Beyond a certain critical
value of $W$, $T$ increases as we increase the disorder strength. It
is clear that the {\em delocalization or the extended states emerge in
  the strong disorder regime}.

In order to study the effect of the location of the line-disorder, we
have plotted the transmission coefficient as a function of the
position of the line-disorder as shown in Fig.~\ref{t}(c). The Fermi
energy is fixed at $E=-0.18$ and the disorder strength is $W=0.5$. The
location of the line is governed by the definition of the width of the
ZGNR as mentioned earlier. A (zigzag) line-disorder moves from the
bottom of the sample to the top, the location of the line number is
given in units of $n$ (see Fig.~\ref{setup}). $T$ is symmetric about
$n=20$ (width being 40A), where a reflection symmetry exists along
$y$-direction.

A comparison between a bulk disordered ZGNR and a set of line-disorder
systems has been studied in Fig.~\ref{t}(d). As mentioned earlier, the
width of the ZGNR is 40A, that is we have a total 40 zigzag lines. The
set of line-disorder samples are taken as follows. We have considered
a single zigzag line, and twenty consecutive zigzag lines from the
bottom (half of the ZGNR width).
\begin{figure}[ht]
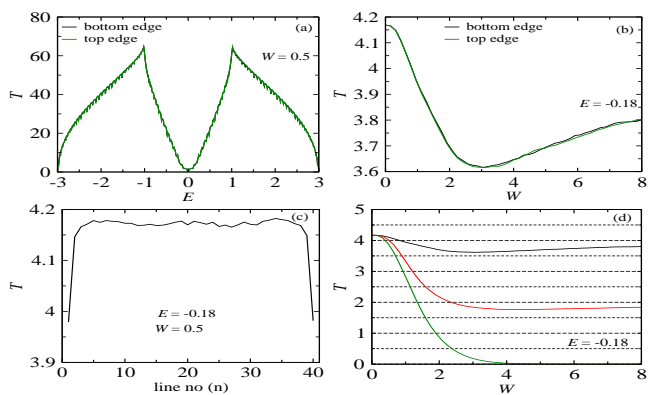

\centering \includegraphics[height=0.14\textwidth,
  width=0.23\textwidth]{fig2a.eps}
\includegraphics[height=0.14\textwidth, width=0.23\textwidth]{fig2b.eps}
\includegraphics[height=0.14\textwidth,
  width=0.23\textwidth]{fig2c.eps}
\includegraphics[height=0.14\textwidth,
  width=0.23\textwidth]{fig2d.eps}
\caption{(Color online). (a) Total transmission coefficient $T$ as a
  function of the Fermi energy for the edge-disorder cases with
  disorder strength $W=0.5$, where the bottom and top edge disordered
  cases are denoted by the black and green lines, respectively. (b)
  $T$ as a function of the strength of disorder for the edge-disorder
  cases at a typical energy $E=-0.18$, where two different colored
  curves represent the identical meaning as given in (a). (c)
  Dependence of total transmission probability on the position of the
  disorder line ($n$) for $E=-0.18$ and $W=0.5$. (d) $T$-$W$
  characteristics at some typical disordered cases where the total
  number of disordered lines are different. Three distinct cases are
  shown where the disorder is introduced at the bottom edge, twenty
  consecutive zigzag lines from the bottom (half of the ZGNR width),
  and, all over the central scattering region. The corresponding
  results are denoted by the black, red and green lines,
  respectively. Here also we choose $E=-0.18$.}
\label{t}
\end{figure}
For convenience we
call these as partially disordered cases. The corresponding
transmission coefficients are denoted by black, and red colors
respectively. The bulk disorder case is denoted by green color. The
complete localization takes place around $W\sim 4$ for the bulk
disorder case, which is the familiar Anderson localization. For the
partially disordered cases, the scenario is completely different as
evident from Fig.~\ref{t}(b). Moreover, if we look at the strong
disorder regime and follow the horizontal dashed lines (drawn to
illustrate the slope of $T$), where $T$ increases with $W$, the rate
of enhancement of $T$ is greater for the single line-disorder than the
half (set of 20 line-disorder) disordered case.
Though the localization/delocalization effect has already been
discussed in a few recent
papers~\cite{zhong-shell-dope,maiti-cpl,yang-physicab} for various
kinds of systems, still we have included this feature for the sake of
completeness and rigour. Zhong and Stocks~\cite{zhong-shell-dope} had
given a proof for this which goes on to explain the anomalous behavior
of $T$ in the following way. We may assume that for the partially
disordered case, the disordered region is coupled with the rest of the
clean (free from disorder but Rashba SO is present) region. If we
assume that there is no coupling between these two regions, then there
will be localized states in the disordered region, while in the clean
region the states will be extended. 
\begin{figure}[ht]
\hfill
\includegraphics[height=0.14\textwidth, width=0.22\textwidth]{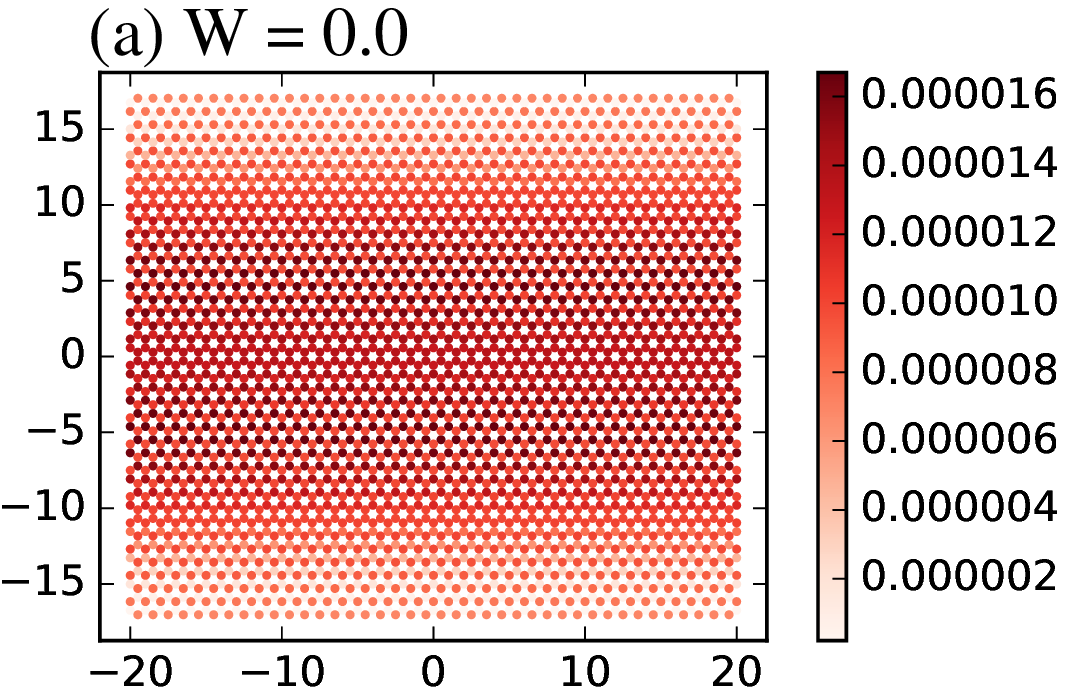}\hfill
\includegraphics[height=0.14\textwidth, width=0.22\textwidth]{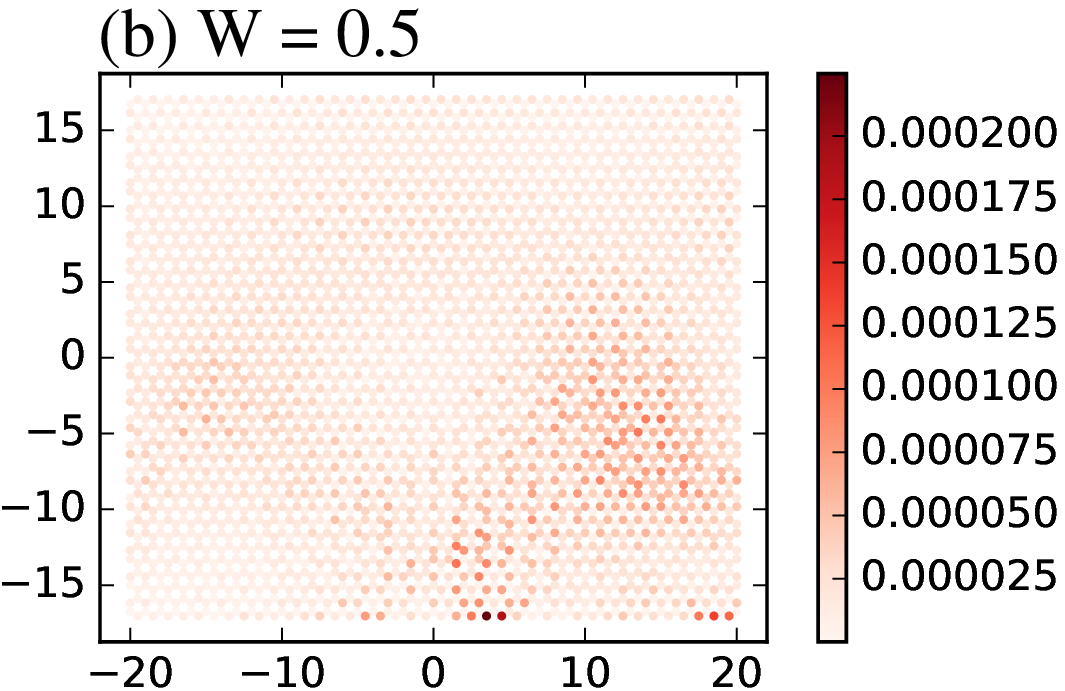}\hfill\vskip 0.1 in
\hfill\includegraphics[height=0.14\textwidth, width=0.22\textwidth]{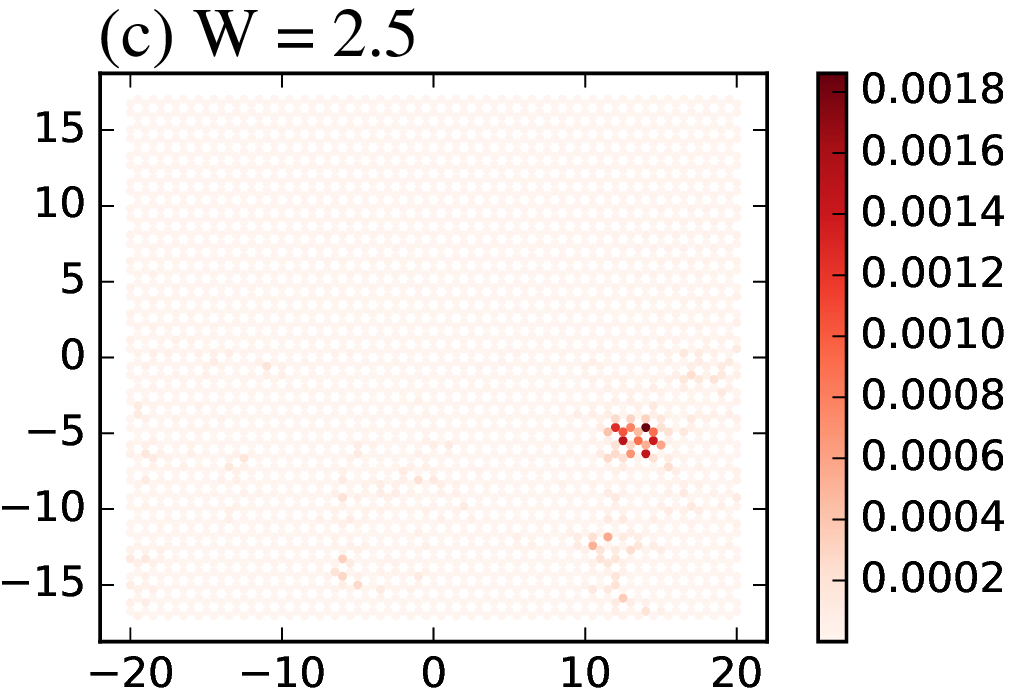}\hfill
\includegraphics[height=0.14\textwidth, width=0.22\textwidth]{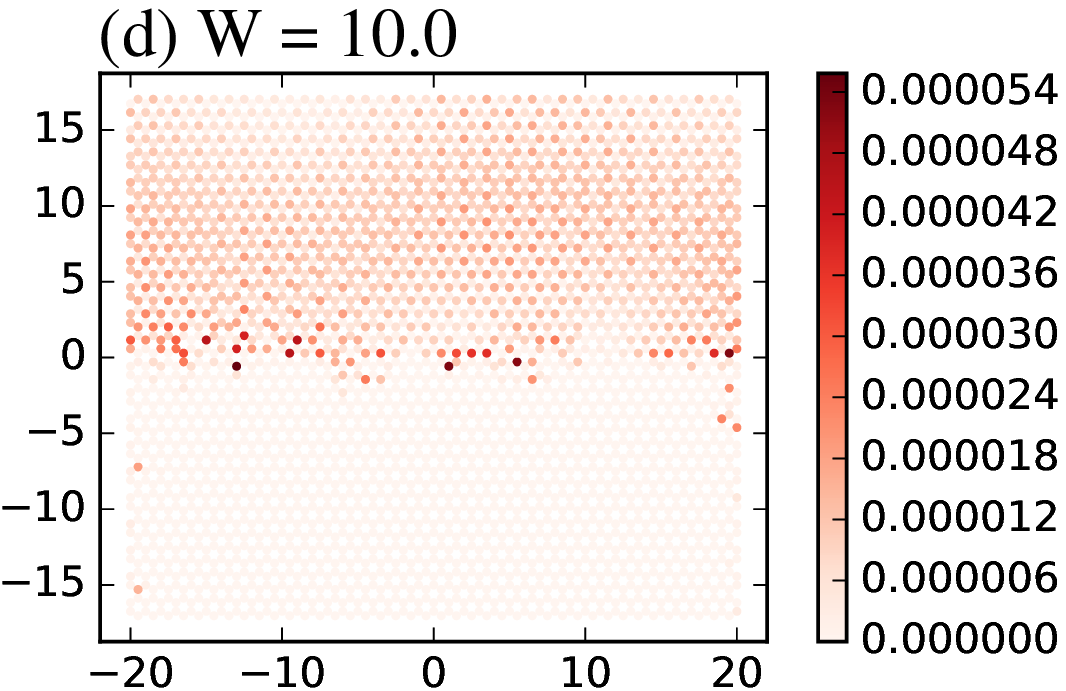}\hfill
\caption{(Color online). LDOS plot for the half disorder
    ZGNR with (a) $W=0.0$, (b) $W=0.5$, (c) $W=2.5$ and (d)
    $W=10.0$.}
\label{ldos}
\end{figure}
Now suppose we switch on the coupling, which in this case will be
the hopping parameter $t$, then there will a competition between the
localized states and the extended states associated with the
disordered and clean regions, respectively.  In the weak disorder
regime, the coupling effect is strong, and thus, the localized states
will affect the electron transport more than that by the extended states,
resulting reduced transmission probability with $W$ in the limit of
weak disorder. On the other hand, the coupling effect gradually
decreases in the strong disorder regime, where the extended states are
less affected by the localized states and as a result, in this limit, the
transmission probability increases with increasing disorder.

The anomalous behaviour mentioned above can also be
  understood from Fig.~\ref{ldos}, where we have plotted the
  space-resolved density of states (LDOS) for three
  different strengths of disorder, including the disorder-free
  case. The partially disordered system is considered as half-disorder ZGNR in
  this case. In the absence of disorder, the LDOS plot
  (Fig.~\ref{ldos}(a)) clearly shows that all the states behave like
  extended states. Now, for $W=0.5$, the localization starts to occur in the disordered region (Fig.~\ref{ldos}(b)). When the
  disorder strength is $W=2.5$ (close to the critical value,
  Fig.~\ref{ldos}(c)), only a few states have non-zero LDOS, while the
  amplitude of other states are vanishingly small. This complete localization
  explains the minimum of the transmission, $T$. Again, for a higher value of disorder,
  namely $W=10$, there is a complete order-disorder phase separation
  as seen from Fig.~\ref{ldos}(d). At this higher value of $W$, all
  the states with non-zero amplitudes are located in the disorder-free region, while
  no such states exist in the disordered region. This discussion also
  clarifies the distinction between `weak' and `strong' disorder regimes. When
  the disorder strength is less than the critical value, where the
  electronic states behave like extended states, we call this
  regime as weak, and beyond the critical value, the regime is strong
  disorder regime.

\subsection{Spin-polarized transmission}

So far, we have studied the total transmission probability for a
variety of line-disorder scenarios as well as for the bulk disordered
case in presence of Rashba SO interaction. Let us now study the
characteristic features of the spin-polarized transmission which is
the central focus of our work. In a pristine GNR,
\begin{figure}[ht]
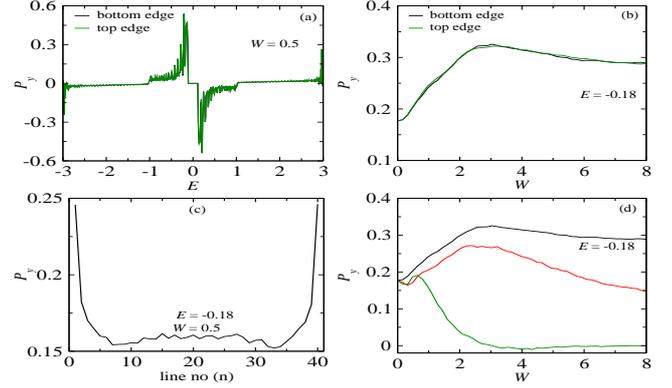

\centering \includegraphics[height=0.14\textwidth,
  width=0.23\textwidth]{fig4a.eps}
\includegraphics[height=0.14\textwidth, width=0.23\textwidth]{fig4b.eps}
\includegraphics[height=0.14\textwidth,
  width=0.23\textwidth]{fig4c.eps}
\includegraphics[height=0.14\textwidth,
  width=0.23\textwidth]{fig4d.eps}
\caption{(Color online). The $y$-component of spin-polarized
  transmission ($P_y$) under different conditions. (a) $P_y$ as a
  function of the Fermi energy for the edge-disorder cases setting
  $W=0.5$. (b) $P_y$-$W$ characteristics for the edge-disorder
  cases. (c) Dependence of $P_y$ on the position of the line-disorder
  ($n$). Here we choose $W=0.5$.  (d) $P_y$ as a function of $W$ for
  some distinct disordered cases, where different colors represent the
  identical meaning as described in Fig.~\ref{t}(d). In the spectra
  (b)-(d) we choose $E=-0.18$.}
\label{py}
\end{figure}
one can have only the $y$-component of the spin-polarized
  transmission due to the longitudinal mirror symmetry of the finite
  width GNR in presence of Rashba SO interaction. However, the
  inclusion of line(s) disorder destroys this symmetry, and as a
  result, the other two components, namely, $P_x$ and $P_z$ start
  contributing in addition to $P_y$~\cite{chico}.

The characteristic features of $P_y$ are shown in Fig.~\ref{py}. In
Fig.~\ref{py}(a), $P_y$ is plotted as a function of the Fermi energy
for the disorder strength $W=0.5$. The behavior of $P_y$ for the top
and bottom line-disorder cases are identical and this feature is
similar to that of the total transmission coefficient as shown in
Fig.~\ref{t}. However, $P_y$ as a function of the disorder strength
shows completely different behavior that of the charge transmission
$T$ (Fig.~\ref{py}(b)), that is, starting with a non-zero value ($P_y
\sim 0.18$ at $W=0$), $P_y$ increases (along the positive direction
i.e., suppressing down spin propagation) up to a certain disorder
strength $W\sim 3.0$ (below this value the Anderson localization
dominates, as observed from Fig.~\ref{ldos}), and then slowly
decreases with increasing $W$. This behavior is peculiar as it implies
that the Anderson localization is beneficial to the spin-polarized
transport for line-disorder ZGNR.

Moreover, as mentioned earlier that {\em due to the finite width of
  the ZGNR the longitudinal mirror symmetry is broken along the
  $y$-axis, even when $W=0$, which results a non-zero $P_y$}. When we
introduce edge-disorder, the system acquires another asymmetric
feature, which in turn aids $P_y$ even in the presence of Anderson
localization. The variation of $P_y$ as a function of the location of
the line-disorder is shown in Fig.~\ref{py}(c). It is symmetric about
$n=20$. The different line-disorder scenarios are shown in
Fig.~\ref{py}(d) and
\begin{figure}[ht]
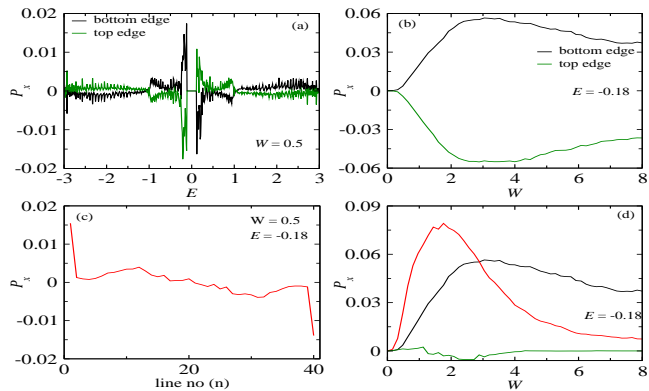

\centering \includegraphics[height=0.14\textwidth,
  width=0.23\textwidth]{fig5a.eps}
\includegraphics[height=0.14\textwidth, width=0.23\textwidth]{fig5b.eps}
\includegraphics[height=0.14\textwidth,
  width=0.23\textwidth]{fig5c.eps}
\includegraphics[height=0.14\textwidth,
  width=0.23\textwidth]{fig5d.eps}
\caption{(Color online). The $x$-component of spin-polarized
  transmission ($P_x$) under same different conditions as described in
  four spectra of Fig.~\ref{py}.}
\label{px}
\end{figure}
colors have the same meaning as mentioned in the context of
Fig.~\ref{t}(d).  For the partially disordered cases, $P_y$ increases
in the weak disorder regime up to a critical disorder strength. This
value of critical disorder is not unique and depends on the number of
line-disorder present in the system. $P_y$ increases beyond this
critical value in the strong disorder regime. For the bulk disorder
case (green curve), $P_y$ shows similar feature as the partially
disordered cases in low disorder regime, whereas in presence of large
disorder $P_y$ drops nearly to zero for $W>4$ due to almost complete
electronic localization.

As the line-disorder is introduced, due to the broken longitudinal
mirror symmetry, finite $P_x$ and $P_z$ are generated. Figure~\ref{px}
shows the characteristic feature of the $x$-component of the
spin-polarized transmission. For the top and bottom line-disorder
cases (denoted by green and black colors, respectively), $P_x$ is
antisymmetric with respect to each other (Fig.~\ref{px}(a)).  $P_x$ is
also antisymmetric as a function of the Fermi energy about $E=0$. The
antisymmetric nature of $P_x$ for the top and bottom edge-disorder
cases is also due to the same reason as given in Fig.~\ref{py}(a) for
the explanation of the symmetric behavior of $P_y$. Figure~\ref{px}(b)
shows the variation of $P_x$ as a function of disorder strength $W$
for the top and bottom line-disorder scenarios. At $W=0$, it is
expected that $P_x$ should be zero due to the longitudinal mirror
symmetry along the $x$-axis, but it remains vanishingly small up to
very low values of $W$. While the inclusion of weak disorder can
destroy the longitudinal mirror symmetry, the localization still
dominates in this regime. $P_x$ again shoots up with increasing
disorder strength, and beyond the critical value $W\sim 3$ (this value
has also been noted from the behavior of $T$ and $P_y$ for
edge-disorder) it decreases with increasing $W$. Moreover, $P_x$ has
equal magnitude and opposite phases for the top and bottom
line-disorder cases, which also make the {\em line-disordered ZGNR not
  only an efficient spin-filter device, but also a switching device.}
This feature can also be verified from the variation of $P_x$ as a
function of the impurity position as shown in Fig.~\ref{px}(c). Here
the phase of $P_x$ continuously changes from being positive to
negative as we move the impurity position from the bottom edge to the
top edge of the ZGNR. Thus, it is possible to tune the magnitude and
determine the phase of $P_x$ by changing the location of the
line-disorder which is undoubtedly an important observation. Finally,
the behavior of $P_x$ as a function of $W$ for the different
disordered cases is shown in Fig.~\ref{px}(d). For the partially
disordered systems, the peculiar behavior (viz, the spin-polarized
transmission enhances in presence of weak disorder, while the reverse
happens for the strong disorder) is again prominent like what we get
in the case of $P_y$ (see Fig.~\ref{py}(d)). For the bulk disorder
ZGNR, $P_x$ shows vanishingly small amplitude and beyond a certain
value of $W$ it completely disappears.

Finally, we have studied the characteristic features of the
$z$-component of spin-polarized transmission and the results are shown
in Fig.~\ref{pz}.  The variation of $P_z$ as a function of the Fermi
energy shows similar nature (Fig.~\ref{pz}(a)) to that of
$P_x$. All the spin-polarized components are antisymmetric
  about $E=0$ as a function of the Fermi energy due to the
  electron-hole symmetry of the system. For the top and bottom
edge-disorder cases,
\begin{figure}[ht]
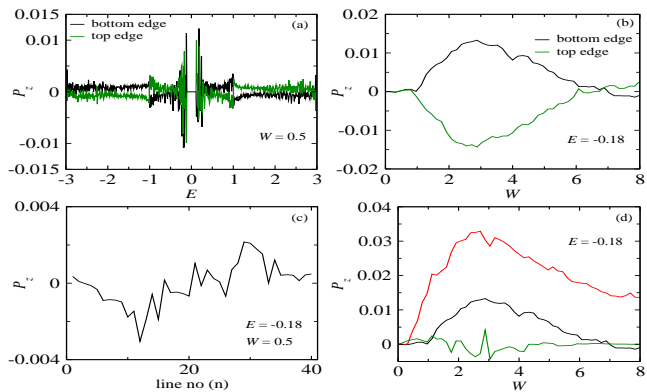

\centering \includegraphics[height=0.14\textwidth,
  width=0.23\textwidth]{fig6a.eps}
\includegraphics[height=0.14\textwidth,
  width=0.23\textwidth]{fig6b.eps}
\includegraphics[height=0.14\textwidth,
  width=0.23\textwidth]{fig6c.eps}
\includegraphics[height=0.14\textwidth,
  width=0.23\textwidth]{fig6d.eps}
\caption{(Color online). The $z$-component of spin-polarized
  transmission ($P_x$) under same different conditions as described in
  four spectra of Fig.~\ref{py}.}
\label{pz}
\end{figure}
$P_z$ has equal and opposite magnitudes (Fig.~\ref{pz}(b)) indicates
the switching property of $P_z$ as we have observed in the case of
$P_x$.  As a function of the location of line-disorder, $P_z$ is
nearly antisymmetric about $n=20$ as shown in Fig.~\ref{pz}(c). The
fluctuations in $P_z$ are owing to the finite size
effects. Figure~\ref{pz}(d) shows the behavior of $P_z$ as a function
of $W$ for different disorder scenarios.  For the half disordered
case, the same peculiar behavior is prominent as seen from the
behaviour of $P_x$ and $P_y$ (Fig.~\ref{px}(d) and Fig.~\ref{py}(d),
respectively).  In presence of bulk disorder, $P_z$ has very small
amplitude, and beyond a certain value of disorder strength, it
completely vanishes.

Though the results presented here have been worked out for
  certain specific parameter values considering a typical system size of
  the ZGNR, all the physical pictures remain valid
\begin{figure}[ht]
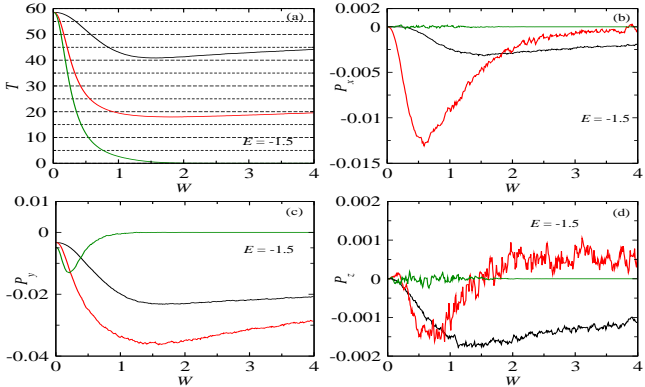

\centering \includegraphics[height=0.14\textwidth,
  width=0.23\textwidth]{fig7a.eps}
\includegraphics[height=0.14\textwidth,
  width=0.23\textwidth]{fig7b.eps}
\includegraphics[height=0.14\textwidth,
  width=0.23\textwidth]{fig7c.eps}
\includegraphics[height=0.14\textwidth,
  width=0.23\textwidth]{fig7d.eps}
\caption{(Color online). (a) $T$, (b) $P_x$, (c) $P_y$ and
    (d) $P_z$ as a function of $W$ for some distinct disordered cases,
    where different colors represent the identical meaning as
    described in Fig.~\ref{t}(d). The Fermi energy is fixed at
    $E=-1.5$.}
\label{diff_set}
\end{figure}
for any other set of parameter values. In support of this, we have
also shown a few results corresponding to a different values of the Fermi energy,
namely, $E=-1.5$ as shown in Fig.~\ref{diff_set}. Here, we have shown
the behavior of $T$, $P_x$, $P_y$ and $P_z$ as a function of disorder
strength $W$. Though the magnitudes of the different spin-polarized
components have lower values compared to those for earlier cases,
the anomalous behavior is still prominent at this particular energy,
which accounts for the robustness of our analysis. Having done a thorough
and extensive numerical study, we observe that this feature is also
robust for any number of disordered lines in a ZGNR. However they are skipped for the sake of brevity. Moreover, it is also important to note that since the
inclusion of the line defect (irrespective of the number of lines), including
the bulk disorder, destroys the longitudinal mirror symmetry of the
system~\cite{chico}, and hence non-zero $P_x$ and $P_z$ will always be generated
along with $P_y$.

\section{Experimental perspective} We propose a possible
  device fabrication technique where a order-disorder structure can
  be realized. Using the `Dip-Pen' nanolithography
  (DPN)~\cite{dip-pen} it is possible to create a selective rough
  surface. Now, if we manage to create such a surface in one half of
  the substrate region, while other half is smooth, then the GNR grown
  on that substrate will acquire the morphology of the substrate. Such
  substrate morphology can also be achieved using other
  nanolithography techniques such as photolithgraphy, X-ray
  lithography with proper masking~\cite{acikgoz} etc.

All the three components of the spin-polarization can be measured by using
a Wien filter and Mott detector (see Ref.~\cite{kisker} for detailed
discussion).

\section{\label{conclusion}Conclusion}

To summarize, in the present work, we have critically investigated the
characteristic features of charge and spin dependent transport
properties of zigzag graphene nanoribbons with line(s) of disorder in
the presence of Rashba SO interaction within a tight-binding framework
based on Green's function formalism. The disordered region is
considered at the edges (top and bottom edges separately), comprising
of a few zigzag lines, and, in order to compare the results of these
spatially non-uniform disordered ZGNRs with a spatially uniform
disordered ZGNR we have considered bulk disordered ZGNR.

The behavior of the charge and spin transport properties for partially
disordered scenario show completely contradicting features to that of
a bulk disordered ZGNR. The charge transmission decreases in the weak
disorder regime due to the Anderson localization, while in the strong
disorder regime the extended electronic states dominate and the charge
transmission increases with increasing disorder strength, suppressing
the effect of electronic localization. All the three components of the
spin-polarized transmission ($P_x$, $P_y$ and $P_z$) demonstrate
completely inverted behavior to that of the charge transport. Our
results predict that for the $y$-component of spin-polarized
transmission, the magnitudes and the phases are same for the top and
bottom line-disorder cases, while for the $x$ and $z$-components the
magnitudes are same but they have opposite phases. The
  symmetry analysis of the $S$-matrix
  elements~\cite{kim-jap,skm1,sudin-sm} also agrees with our findings.
Since the presence of line-disorder can generate all the three
components of the spin-polarized transmission, a line-disorder ZGNR
can be implemented as an efficient spin-filter device. Moreover, by
studying the effect of the location of line impurity, the system can
be used to tune the magnitude as well as the phase of $P_x$. Since the
$x$ and $z$-components have the same magnitude but opposite phases for
the top and bottom line-disorder cases, the system can be utilized as
a switching device.

\setcounter{secnumdepth}{0}

\section{ACKNOWLEDGMENT}

SG thanks Dr. A. Gayen for fruitful discussion on the experimental
perspective. SB thanks SERB, India for financial support under the
grant F. No: EMR/2015/001039.

\end{document}